\documentclass[aps,pre,reprint,floatfix,superscriptaddress]{revtex4-1}
\usepackage{graphicx,lipsum}
\usepackage[latin1]{inputenc}
\usepackage{amsmath}
\usepackage{amsfonts}
\usepackage{amssymb}

\begin{document}

\date{\today}

\title{Tunable Order of Helically Confined Charges}

\author{Ansgar~Siemens}
\email{asiemens@physnet.uni-hamburg.de}
\affiliation{Zentrum f\"ur Optische Quantentechnologien, Fachbereich Physik, Universit\"at Hamburg, Luruper Chaussee 149, 22761 Hamburg Germany} 
\author{Peter~Schmelcher}
\email{pschmelc@physnet.uni-hamburg.de}
\affiliation{Hamburg Center for Ultrafast Imaging, Universit\"at Hamburg, Luruper Chaussee 149, 22761 Hamburg Germany}
\affiliation{Zentrum f\"ur Optische Quantentechnologien, Fachbereich Physik, Universit\"at Hamburg, Luruper Chaussee 149, 22761 Hamburg Germany}

\begin{abstract}

\noindent We investigate a system of equally charged Coulomb-interacting particles confined to a toroidal helix in the presence of an external electric field. 
Due to the confinement, the particles experience an effective interaction that oscillates with the particle distance and allows for the existence of stable bound states, despite the purely repulsive character of the Coulomb interaction. 
We design an order parameter to classify these bound states and use it to identify a structural crossover of the particle order, occurring when the electric field strength is varied. 
Amorphous particle configurations for a vanishing electric field and crystalline order in the regime of a strong electric field are observed. 
We study the impact of parameter variations on the particle order and conclude that the crossover occurs for a wide range of parameter values and even holds for different helical systems. 

\end{abstract}

\maketitle

\section{Introduction}

One-dimensional (1D) interacting many-body systems are a field of steadily growing interest, in particular due to the fact that the properties of these 1D structures can be drastically different from those of bulk materials \cite{lan2011,samuelson2004}.
This renders them interesting candidates for e.g. nano-electronics and photonic applications \cite{avouris2006,hatton2008,avouris2008}. 
Prominent approaches for confining particles to 1D space include carbon nanotubes (CNT) \cite{ijima1993,shaikjee2012,lau2006} as 1D nanowires, the trapping of particles in evanescent fields of thin nanofibers \cite{garcia-fernandez2011,vetsch2012,reitz2012,vetsch2010}, and biological compounds such as DNA molecules \cite{tran2000,yu2001}.
The last example is especially interesting since the 1D nanowires are not straight but bent into a helix.
The helical structure offers several advantages, such as increased robustness with regards to deformations \cite{marko1994,kornyshev2007}, which makes them a desirable class of systems.
One-dimensional structures with helical geometry are not limited to biological compounds but are prominent in artificial nanostructures and photonics \cite{lau2006,shaikjee2012,prinz2000,reitz2012,schmidt2001,bhattacharya2007}.
Experimental preparatory techniques have succeeded in designing helical structures with diameters as small as $10\ nm$ \cite{ivanov1994,shaikjee2012}.

Charged particles confined to a helical structure can give rise to interesting properties, such as the optical activity of chiral molecules \cite{drude1906,maki1996}.
However, many approaches to the physics of helical systems are based on approximations with continuous charge densities or noninteracting particles. 
In recent years, a series of theoretical works have demonstrated that a fascinating behavior can emerge when long-range interactions between helically confined particles are taken into account. \cite{schmelcher2011,kibis1992,pedersen2014,zampetaki2013,zampetaki2015,zampetaki2015a,zampetaki2017,zampetaki2018,plettenberg2017,pedersen2016a,pedersen2016}. 
It was shown that the Coulomb interaction, together with the constraining forces of the helix can form an effective 1D interaction potential that oscillates with the particle distance on the helix \cite{schmelcher2011,kibis1992}.
Depending on the helix geometry, this potential can possess several minima, at which the Coulomb forces are exactly canceled by the constraining forces. 
This allows the particles to `condense' into stable lattice-like 1D particle chains on the helix, even when their interactions are purely repulsive.

As a direct consequence of the oscillating effective interactions, interacting particles on inhomogeneous helices were shown to exhibit interesting dynamics, such as the binding or dissociation of particles by scattering at such an inhomogeneity \cite{zampetaki2013}.
Investigations of interacting particles on a toroidal helix showed that the band structure and dispersion of the phonons can be controlled by the helix radius \cite{zampetaki2015}.
For the toroidal helix, a critical radius was found, at which the oscillations of individual particles decouple, and excitations are prevented from dispersing through the helical system \cite{zampetaki2015,zampetaki2015a}. 

The unique properties arising from the oscillating effective interactions are not limited to the structure and dynamics of charged particles. 
The overall mechanical behavior of the helix is equally affected.
This can, for example, be observed in the unusual electrostatic resistance to a bending of the helix \cite{zampetaki2018}: 
When interactions between the particles are considered, it was found, that for a fixed particle density, the system switches periodically between favoring and resisting the bending when the length of the helix is varied. 
Additionally, 
for large helix radii, the system's ground state (GS) was found to drastically change for just slight variations in the bending, resulting in a discontinuous GS bending response.

The above examples demonstrate the unique physics occurring when long-range interactions between helically confined particles are considered. 
A common denominator of these investigations is the dependence of the unique properties on the helix geometry. 
Since variations - e.g. of the helix radius - are difficult to realize in experimental setups, the question arises whether a similar control can be achieved in a different way, such as by the application of external electric fields.
A first step in that direction was taken in Ref. \cite{plettenberg2017}, where time-dependent electric fields were proposed to realize state transfers between arbitrary equilibrium configurations in a helical system with two charged particles.

In the above spirit, we study here the influence of an electric field on the static properties of charged particles on a toroidal helix.
We identify two distinct phases of order that depend on the external field strength: 
an amorphous-like phase, that persists for weak electric fields, and a phase with crystalline order that is adopted in the presence of strong electric fields.
We demonstrate the possibility of continuously switching between the phases by varying the electric field strength.
Furthermore, we verify the existence of the observed phases for a large parameter space.

This work is structured as follows.
Section \ref{Setup} describes the confinement of charged particles to a toroidal helix and the resulting effective interactions in detail.
In section \ref{example}, we then show the structural phase transition for an example system. 
We start by demonstrating amorphous ordering for individual particle configurations.
Subsequently, an order parameter is designed and the disorder of the system is classified.
Then, we demonstrate how a continuous transition from amorphous to lattice ordering can be induced by tuning the electric field strength.
In section \ref{Generalization} we demonstrate the generality of our results for a wide parameter range. 
Our conclusions and outlook are provided in section \ref{summary}.

\section{Charged Particles on a Toroidal Helix}
\label{Setup}

We consider a system of $N$ equally charged particles confined to a 1D path defined by a parametric function $\textbf{r}(u):\mathbb{R}\rightarrow\mathbb{R}^3$.
The particles are subject to a gradient force $\textbf{F}_E=q\textbf{E}$ in form of an external electric field, and interact via Coulomb forces $\textbf{F}_C=\lambda\textbf{e}_{ij} /|\textbf{r}(u_i)-\textbf{r}(u_j)|^2$, where $\lambda=q^2/4\pi\epsilon_0$ is the coupling constant, and $u_i$ and $u_j$ are the positions of the interacting particles in parametric coordinates.
Since the particles are only allowed to move along the 1D path $\textbf{r}(u)$, they additionally experience confining forces.
The confining forces cancel all forces acting perpendicular to the path, resulting in an effective force parallel to the tangential vector $\partial_u\textbf{r}(u)$.
Mathematically, the consideration of confining forces corresponds to a projection of forces on the parametric curve.
With 
\begin{equation}
proj\left(\textbf{a},\textbf{b}\right):=\left(\textbf{a}\cdot\textbf{b}\right)\cdot\dfrac{\textbf{b}}{||\textbf{b}||}\nonumber
\end{equation}
the effective force on a particle can be written as
\begin{equation}
\textbf{F}_{eff}^{(i)}
=proj\left(q\textbf{E}+\sum_{i\neq j}^N\dfrac{\lambda\ \textbf{e}_{ij}}{|\textbf{r}(u_i)-\textbf{r}(u_j)|^2}\ ,\ \dfrac{d\textbf{r}(u_i)}{du_i}\right)
\end{equation}
The effective force on a particle vanishes if the sum of all forces is perpendicular to the confining path.
Despite repulsive interactions between the particles, the geometry of the path can thereby allow for the existence of equilibrium states where the effective forces on all particles vanish.
These stable equilibrium configurations correspond to minima in the potential energy, which is given by
\begin{align}
V_{tot} = V_E+V_C =  \sum_{i=1}^{N}q\textbf{E}\cdot\textbf{r}(u_i)+\sum_{i<j}^{N} \frac{\lambda}{|\textbf{r}(u_i)-\textbf{r}(u_j)|}
\label{eq:2}
\end{align}
Eq. \ref{eq:2} already accounts for the effects of the confinement by only allowing for particle positions on the parametric curve $\textbf{r}(u_i)$.

\begin{figure}
\includegraphics[width=\columnwidth]{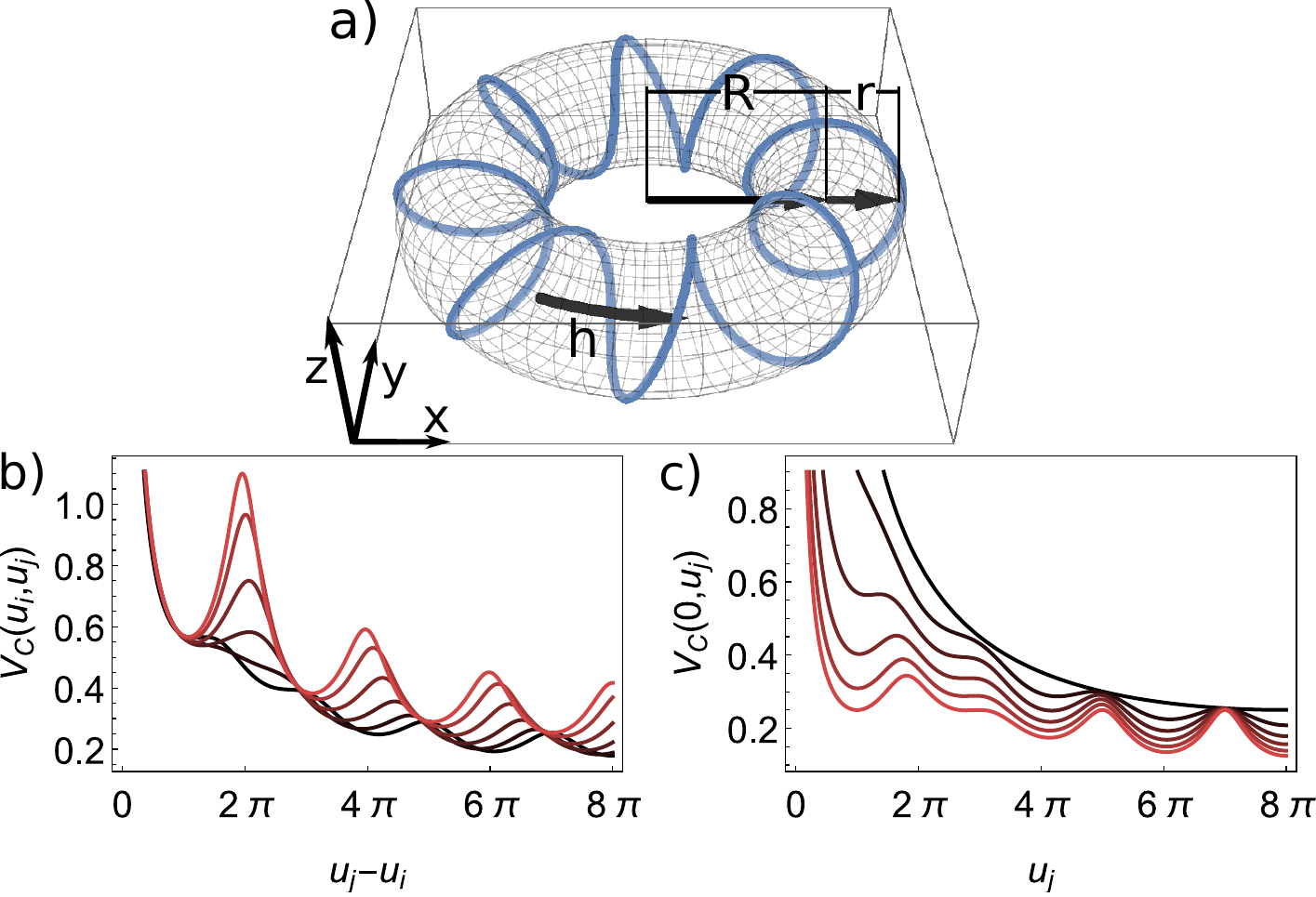}
\caption{\label{figure1}(a) A toroidal helix with $M=8$ windings, for the parameter values $r/h=1.6/\pi$, $R=2$. (b) The Coulomb potential $V_C(u_i,u_j)$ for the same parameters as (a), and different fixed values of $0\leq u_i\leq\pi$. The position $u_j=\pi M=8\pi$ corresponds to particle positions on opposite sides of the torus. Coloring varies from red ($u_i=\pi$) to black ($u_i=0$). (c) The Coulomb potential $V_C(0,u_j)$ for the same parameters as (a), and different helix radii $0\leq r\leq R$. Coloring varies from black for $r=0$ to red for $r=R$. }
\end{figure}

The effects of confining forces are particularly relevant for paths with nontrivial geometry in the form of non-vanishing curvature.
A simple class of systems satisfying this demand - while at the same time being common enough to occur in nature - are helical systems.
Here, we investigate the properties of equilibrium configurations on a toroidal helix. 
An example of such a system is visualized in Fig. \ref{figure1}(a). 
The particle positions on a toroidal helix are given by the following parametrization in Euclidean space:
\begin{equation}
\textbf{r}(u_i):=\left(
\begin{array}{c}
\left(R+r\cos(u_i)\right)\cos(u_i/M) \\
\left(R+r\cos(u_i)\right)\sin(u_i/M) \\
r\sin(u_i)
\end{array}
\right),
\begin{array}{c}
u_i\in [0,2\pi M]\\
i\in[1...N]
\end{array}
\label{eq:1}
\end{equation}
where $R$ is the torus radius, $r$ is the helix radius, and $M$ the number of windings. 
Since we want the helix to close after circling around the torus exactly once, the parameters must satisfy the relation $Mh=2\pi R$, where $h$ is the helix pitch. 
The parametric coordinate $u_i$ of the i-th particle can be interpreted as an angle. 
If $u_i$ changes by an amount of $2\pi$, the position on the helix changes by exactly one winding. 
Since the toroidal helix has $M$ windings, $\textbf{r}(u_i)$ is invariant under translations by $u_i\rightarrow u_i+2\pi M$. 
For easier comparison between systems with different parameters, we introduce the filling factor $v=N/M$ as the ratio of particles per winding.

\paragraph*{Coulomb Potential}
Since the confinement can drastically change the underlying potential landscape, we will now discuss the effects of the confinement on the two sums of Eq. \ref{eq:2} individually.
For the Coulomb potential,
already the contribution of two particles shows an immense complexity compared to the Coulomb interaction of `free' particles. 
When written as a direct function of the parametric coordinates $u_i$ and $u_j$, the two-particle Coulomb potential $V_C^{ij}:=V_C(u_i,u_j)$ of the particles $i$ and $j$ is given by:
\begin{widetext}
\begin{equation}
V_{C}^{ij}= 2\lambda\left[\sin^2\left(\dfrac{u_i-u_j}{2M}\right)\left(R+r \cos(u_i)\right)\left(R+r \cos(u_j)\right)+r^2\sin^2\left(\dfrac{u_i-u_j}{2}\right)\right]^{-\dfrac{1}{2}}
\label{eq:3}
\end{equation}
\end{widetext}
This potential directly depends on the helix parameters and can - depending on the helix geometry - possess several minima. 
Minima in the Coulomb potential correspond to positions, where the particles cannot move further along the helix, without decreasing their (3D) distance to each other - for example by trapping each other on opposite sides of a helix winding. 
The general behavior of Eq. \ref{eq:3} is indicated by Fig. \ref{figure1}(b) which shows multiple cross-sections of the potential for various fixed values of $u_i$.
Contrary to the Coulomb interaction of `free' particles, the effective Coulomb interaction of particles on a toroidal helix cannot be described by the relative particle distance alone. 
As a consequence, the cross-sections of $V_{tot}$ are different for different positions of $u_i$. 
This also implies that both particles can experience slightly different potential wells, and compete for minimization of the total potential energy.
Note that the above-described behavior is a direct consequence of the toroidal helix being bent since in the case of a homogeneous helix the effective interaction can be described using only the relative particle distance. 
The influence of the helix radius on the effective Coulomb potential is shown in Fig. \ref{figure1}(c). 
The figure shows the change of a single cross-section for variations of $r$ in the range $0\leq r\leq R$. 
With decreasing $r$, the stability of the minima decreases. 
For small values of $r$, some of the equilibria become unstable. 
In the limit of $r\rightarrow0$, the oscillations disappear, leaving only a single stable minimum.
In this minimum, both particles are positioned on opposite sides of the toroidal helix with a distance of $\Delta_{ij}=\pi M$.

\paragraph*{Electric Field}
The potential energy contribution of the electric field $V_E$ is similarly influenced by the confining forces.
In this paper, we consider an electric field in z-direction. 
The potential energy $V_E$ then simplifies to:
\begin{equation}
V_E=\sum_{i=1}^N qEr \sin(u_i)
\label{eq:Ve}
\end{equation}
Similar to the Coulomb potential, $V_E$ also oscillates with the particle positions.
Both interactions therefore support the formation of equilibrium configurations. 
However, the characteristic length scales of the oscillations of $V_E$ and $V_C$ are in general different. 
From this, one may already expect fundamentally different behavior for the regimes of dominating Coulomb interaction and dominating electric field. 

From the above discussion, it is easy to see that the potential landscape of a system with $N$ particles can become quite complex since it consists of $(N-1)!$ sums of functions like Eq. \ref{eq:3}, in addition to the potential $V_E$ given by Eq. \ref{eq:Ve}.
Already systems with only a few particles can support a plethora of minima \cite{schmelcher2011} that in general can only be found within numerical calculations. 
With this, the tasks of finding the equilibria of systems with large particle numbers can quickly become computationally expensive.

In the following, we will describe the structural phase transition of the equilibria, occurring when the electric field strength is varied. 
Our results are given in dimensionless parameters, where energies are measured in units of $\lambda/\alpha = q^2/4\pi\alpha\epsilon_0$, and - due to the scale invariance - distances can be scaled by the constant $\alpha=\pi/2h$. 
For an initial overview over the effective behavior, we use the following parameter values: $M=8$, $N=10$, $r=0.8$, $R=Mh/2\pi=2$.

A final remark on our computational approach is in order: 
The minima of $V_{tot}$ for $E=0$ are obtained with a quasi-Newton method \cite{scales1985,nocedal2006}. 
For different electric field strength, the minima were obtained using an interior-point method \cite{boyd2004}, by step-wise varying $E$ and calculating the new minima while using the minima of the previous step as an initial guess.

\section{Structural crossover}
\label{example}

In this section, we investigate the equilibrium configurations of charged particles on the toroidal helix and show how a transition from amorphous-like to crystalline particle ordering can be induced by varying the external electric field strength.
While this structural crossover is a general effect that can be observed for a wide parameter regime, the specific equilibria can be influenced by the helix parameters.
Consequently, the results shown in this section will contain some parameter-specific features.
A generalization of the results to different parameter regimes is discussed in section \ref{Generalization}.

\subsection{The Minima of the Helical Coulomb Potential}

\begin{figure}
\includegraphics[width=\columnwidth]{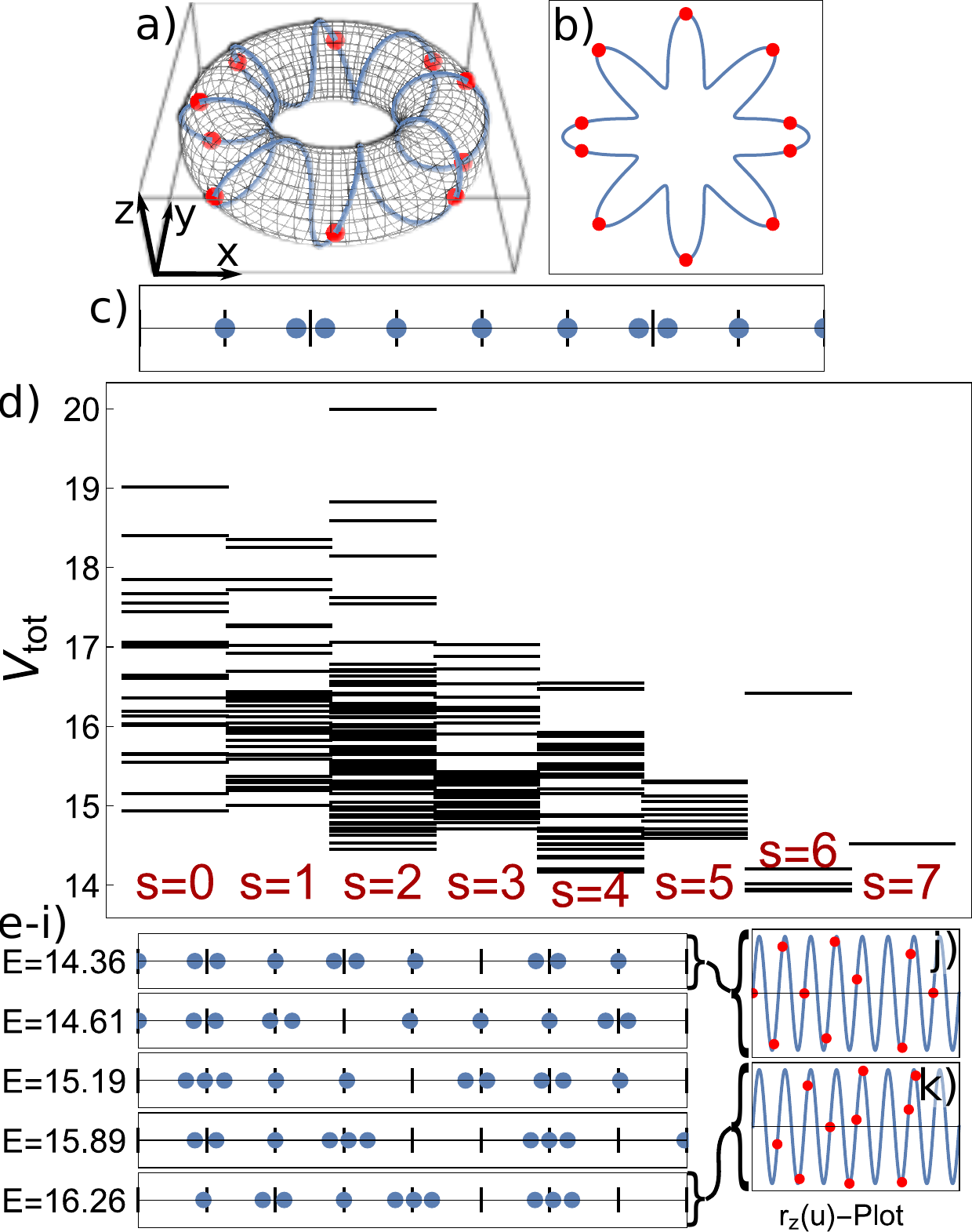}
\caption{\label{figure2}(a)-(c) Visualizations of the ground state: (a) 3D view, (b) xy-projection, and (c) parametric coordinates. Vertical lines in (c) indicate the outermost part of a winding. (d) Energies of stable states sorted by the number of singlets $s$. (e)-(i) Example minima in parametric coordinates, sorted by their energy. Vertical lines again indicate the outermost part of each winding. (j)-(k) Different visualizations of the equilibria of (e) and (i), showing the z-component of the particle positions $r_z(u)=r\sin(u)$ as a function of the parametric coordinate. For emphasis, the possible particle positions are marked by the blue line. }
\end{figure}

We start with the case of a vanishing external electric field. 
In this case, the potential landscape simplifies to $V_{tot}=V_C$. 
Here, we want to convey a basic intuition for the equilibrium particle configurations in this regime. 

\paragraph*{The Ground State}

We start by examining the GS of the system, visualized in Figs. \ref{figure2}(a)-(c). 
For $E=0$, the particles minimize their energy by maximizing their distances. 
Our system is small enough such that all particles interact significantly with each other, and positions close to the center of the torus are avoided.
If we had as many particles as windings (filling factor $v=1$), in the GS, each particle would occupy the outermost point of a winding.
However, since we have $10$ particles and only $8$ windings, some of the helix windings are occupied by several particles.
When two particles occupy the same winding \footnote{windings are separated at positions closest to the center of the torus, i.e. odd multiples of $\pi$ in parametric coordinates}, they want to align at a distance corresponding to the first minimum of the two-particle Coulomb interaction.
Slight deviations from this distance are caused by interactions with neighboring particles.

\paragraph*{Nomenclature}
From the top-view of our system in Fig. \ref{figure2}(b) it is intuitive that the energy of any configuration will increase if a particle moves closer towards the center of the torus.
We will now use this effect to introduce some helpful nomenclature. 
We split the toroidal helical path into $M$ windings and characterize the minima by the distribution of particles onto these windings.
These distributions will be called \textit{x-lets}:
a winding with only a single particle is called a $\textit{singlet}$, one with two particles a $\textit{doublet}$, and so on.
By this definition, the GS depicted in Figs. \ref{figure2}(a)-(c) consists of 6 singlets and 2 doublets.

\paragraph*{Excited States}
An overview of the excited states is given by Fig. \ref{figure2}(d).
It shows the energy of all minima sorted by their number of singlets $s$.
Each minimum is represented by a black horizontal line.
The lowest energy or ground state can be found in the column for states with $s=6$ singlets.
In total 710 distinct minima were found in the potential landscape for $E=0$.
The minima seem to be equally distributed over a large energy range.
We observe a decrease of the occupied energy range for an increasing number of participating singlets.
Example visualizations of excited states are shown in Fig. \ref{figure2}(e)-(i) together with their corresponding energy.
The particle positions in Figs. \ref{figure2}(e)-(i) follow a simple pattern: the particles try to accumulate around the outermost part of a winding.
If they share a winding with other particles they somewhat deviate from this lattice order, creating staggered particle configurations.
The distance between particles that share a winding is approximately the same for doublets and triplets, and corresponds to about $30\%$ of the length of a winding - a significant deviation from a crystalline lattice ordering.

The distances between particles in doublets and triplets are visualized in Fig. \ref{figure2} (j) and (k). 
They respectively show the same equilibria as Fig. \ref{figure2} (e) and (i) and additionally include the z-position of each particle. 
To guide the eye, the z-components of the parametric curve $r_z(u)=r \sin(u)$ is represented by the blue curve. 
We can see that for example the particles in doublets prefer to align almost on opposite sides of the winding to maximize their distance. 

For our chosen parameters, all minima follow the same pattern as the five example excited states in Fig. \ref{figure2}(e)-(i):
the total particle number $N$ is split up into singlets, doublets, triplets and in some cases even quadruplets, which are then distributed among the windings.
Therefore, every minimum is uniquely characterized by a specific sequence of x-lets.

\subsection{Classifying the Particle Order}

Now, we design an order parameter to classify the particle order of every equilibrium particle configuration in a mathematically more rigorous way.
We start by considering how such an ordered particle configuration might look like.
Any quantifiable order that can be observed for most of the minima will likely be linked to a symmetry of the confining manifold.
The toroidal helix possesses a discrete rotational symmetry. 
It is invariant under a rotation of $2\pi n/M$ with $n\in\mathbb{N}$ around its center.
In our parametric coordinates $u_i$ this corresponds to a discrete translation symmetry.
We have already seen hints for this translation symmetry in the discussion above. 
From Eq. \ref{eq:Ve}, we can see that this symmetry holds even in the presence of an electric field in z-direction.
Our order parameter should therefore also reflect this symmetry.
From Eq. \ref{eq:1} and the definition of $V_E$, we can also see that the symmetry is broken if the field has a component perpendicular to the z-direction, which is why this case is not considered in this work.
With this in mind, we define our order parameter $\Theta$:
\begin{align}
\Theta=1-\frac{1}{\pi}\sqrt{\frac{\sum_i(d_i-\mu)^2}{N}} \quad\mathrm{with}\quad  \mu=\dfrac{1}{N}\sum_i^N d_i
\label{eq:4}
\end{align}

where $d_i = Mod\left(u_i,2\pi\right)-Mod\left(u_{i+1},2\pi\right)$ are the truncated next-neighbor(tNN) distances and $\mu$ the mean of all $d_i$. 
The $d_i$ can be interpreted as a relative deviation of the neighboring particle positions within their winding.
The second summand of $\Theta$ in Eq. \ref{eq:4} is simply the normalized standard deviation of the tNN distances.

As an order parameter, $\Theta$ is in the range $[0,1]$.
$\Theta$ is $1$ if all $d_i$ are the same and the particles show a lattice symmetry.
$\Theta$ decreases with increasing deviation from this translation symmetry.
The lower the value of $\Theta$, the greater the disorder.

\begin{figure}
\includegraphics[width=\columnwidth]{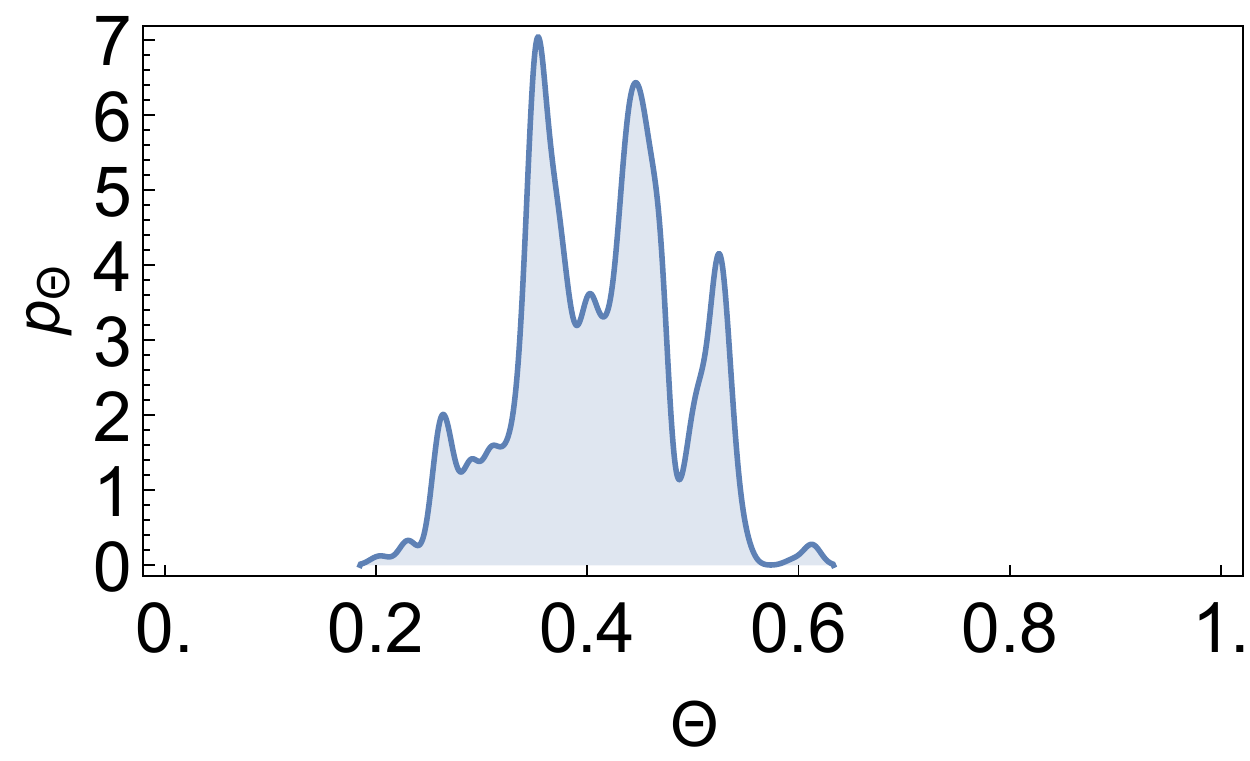}
\caption{\label{figure4}Probability density $p_{\Theta}$ for finding a minimum with a specific order $\Theta$ for a system with $M=8$, $r=0.8$, $h=\pi/2$, $R=2$ and $N=10$.}
\end{figure}

We can now use this order parameter to characterize the order of the equilibria. 
From Fig. \ref{figure2}(b) we can see that the particles in the GS are mostly aligned at the outermost point of each winding.
With the exception of the two doublets, the particles already possess a translation symmetry.
However, the two doublets distort six of the ten tNN-distances from perfect lattice ordering, resulting in $\Theta_{GS}=0.613$.
The order of the minima shown in Fig. \ref{figure2}(e)-(i) is $0.535$, $0.517$, $0.377$, $0.453$, and $0.342$ respectively.
In this fashion, we can calculate the order parameter for every minimum.
The results are shown in Fig. \ref{figure4} as a probability density $p_{\Theta}$. 
The order parameter $\Theta$ of a minimum is (approximately) proportional to the number of singlets $s$. 
The large peak around $\Theta\sim0.53$ consists entirely of minima with $4\leq s\leq6$.
The peaks around $\Theta\sim0.45$ and $0.35$ consist of minima with $2\leq s\leq4$ and $0\leq s\leq3$ respectively.
As we will see, $p_{\Theta}$ will evolve as a bulk when the electric field strength is increased.
For our purposes, the statistics of $p_{\Theta}$ is well described by the mean value $\overline{p_{\Theta}}$ and the standard deviation $\sigma(p_{\Theta})$.
For the curve of Fig. \ref{figure4} we get $\overline{p_{\Theta}}=0.420$ and $\sigma(p_{\Theta})=0.071$.

\subsection{Forcing Crystalline Ordering}

We will now investigate the effect of a static external electric field on the equilibrium particle positions.
From Eq. \ref{eq:Ve}, we know that the electric field creates an additional periodic modulation of the potential landscape. 
The competition between the electric field and the Coulomb forces can induce complex behavior, such as the creation or annihilation of minima when the field strength is varied. 
For the evaluation, the minima for $E>0$ are obtained by tracing the minima found for $E=0$ while the electric field is increased adiabatically, i.e. we do not actively search for newly created minima, which are only stabilized by the electric field.
In practice, we varied the electric field in steps of $\Delta E=0.002$ and calculated the new equilibrium positions using the equilibria of the previous step as initial guesses.

\begin{figure*}
\includegraphics[width=\linewidth]{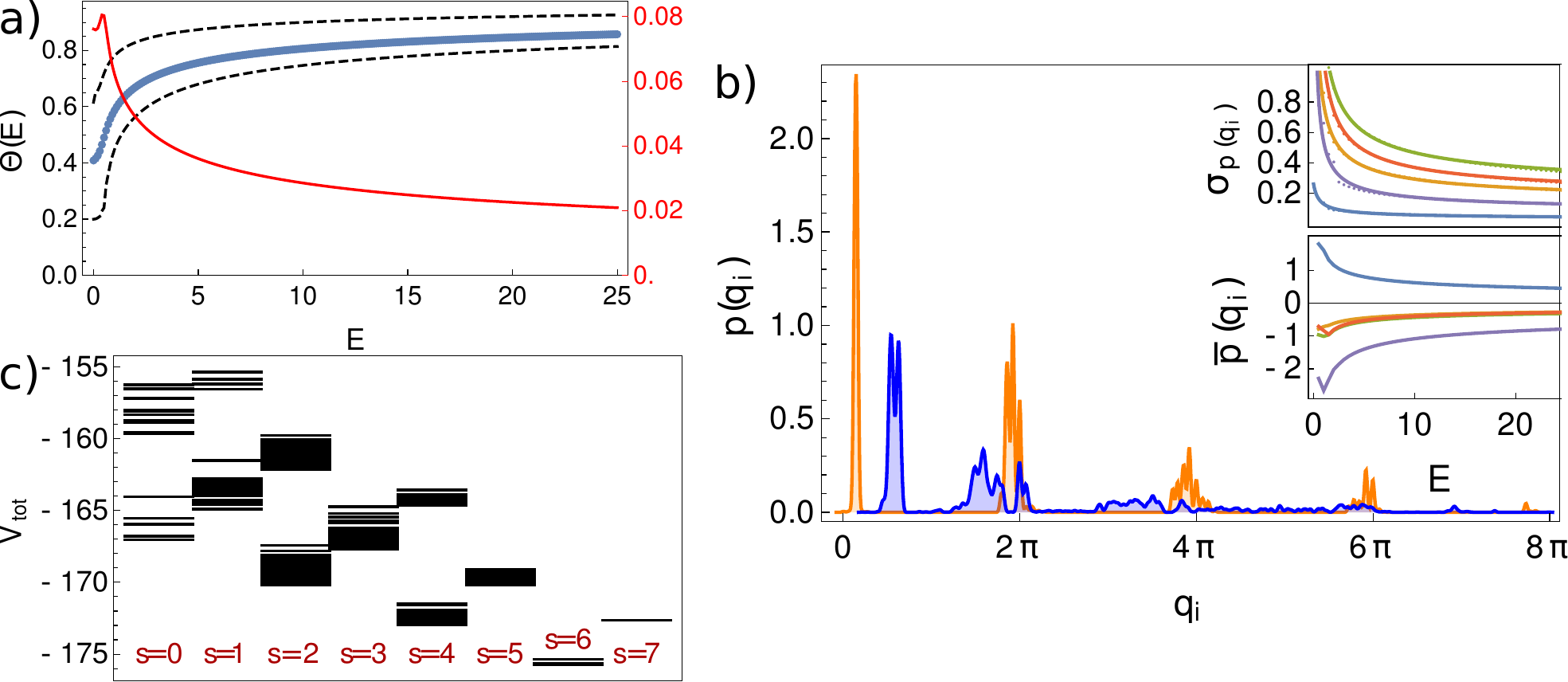}
\caption{\label{figure5}(a) Evolution of the order parameter $\Theta$ when the electric field $E$ is increased adiabatically; with the mean value $\overline{p_{\Theta}}$ (blue), the minimum and maximum $Min(p_{\Theta})$/$Max(p_{\Theta})$ (black), and the variance $\sigma(p_{\Theta}(E))$ (red). Note the different scale of the variance. (b) Normalized statistics of nearest neighbor distances for $E=0$(blue) and $E=25$(orange). The two insets correspond to the variance $\sigma_{p(q_i)}$ and mean (modulo $2\pi$) $\overline{p}(q_i)$ of individual peaks in the statistics as a function of $E$. The colors of the five peaks (sorted by increasing $q_i$) are blue, yellow, green, red and purple. (c) Energies of equilibrium configurations for $E=25$, sorted by their number of singlets $s$.}
\end{figure*}

\paragraph*{Order Parameter Evolution}
The evolution of $p_{\Theta}(E)$ for an adiabatic increase of $E$ can be seen in Fig. \ref{figure5}(a).
The information is split up into the mean value $\overline{p_{\Theta}}(E)$ (blue), the minimum $min(p_{\Theta}(E))$ and maximum $max(p_{\Theta}(E))$ (black), and the standard deviation $\sigma(p_{\Theta}(E))$ (red).
Note the different scale of the standard deviation.
Close to $E=0$, there is a small range where $\overline{p_{\Theta}}(E)$ changes very little, while the variance increases.
In this field range, most of the complex changes in the potential landscape, including the annihilation of minima, take place.
For our chosen parameters, about $\sim16.5 \%$ of all minima are annihilated in this phase.
When a minimum is annihilated, there will be a sudden jump in the tNN distances, corresponding to particles changing their winding.
This sudden change of particle positions is the reason for the increase of $\sigma(p_{\Theta})$ in the low field regime.
The specific bifurcation scenarios by which these annihilations take place depend strongly on the chosen system parameters and can quickly get quite complex \cite{plettenberg2017}.
For larger fields, $\overline{p_{\Theta}}(E)$ increases with $E$ while the variance decreases, indicating a transition to minima with lattice order.
In this regime, variations in the electric field only adjust the particle positions while their distribution among the windings persists.
From $min(p_{\Theta}(E))$ we can see that a large enough field can impose order in the system - independent of the initial particle configuration.

\paragraph*{Relative Particle Positions}
The reason why an increase of the electric field strength causes an increase in our order parameter can be understood from the behavior of individual particles during the transition. 
For this purpose, we consider the nearest-neighbor (NN) distances $q_i:=u_i-u_{i+1}$ of the particles.
Fig. \ref{figure5}(b) shows a statistic of the occurring NN distances for $E=0$ (blue) and $E=25$ (orange). 
For $E=0$, we see a relatively sharp peak at a distance of slightly less than $\pi$, very close to the first minimum in the Coulomb potential.
It is caused by the particles with a NN in the same winding i.e. the doublets, triplets and so on.
It is also possible that the NN particle is in a neighboring winding.
Therefore, we also see NN distances around the value of $2\pi$.
There are also smaller probabilities for finding a NN at distances around $4\pi$, $6\pi$, and $8\pi$.
They correspond to the cases with $1$,$2$, and $3$ empty windings between the NN particles.

The evolution of this statistics with variations of the electric field strength is shown in the insets of Fig. \ref{figure5}(b).
In both insets, the information is split up into $5$ curves, each representing the evolution of a peak in the probability density.
The upper inset shows that the variance of each peak decreases with increasing $E$, while the lower inset describes a continuous shift of the mean position of each peak to multiples of $2\pi$ when $E$ is increased.
For large electric fields, the result are sharp peaks in the distribution $p(q_i)$ at distances of $2\pi n$, $n\in\mathbb{N}_0$ (see Fig. \ref{figure5}(b)(orange)), corresponding to a lattice ordering with a lattice constant of $2\pi$.
The dominating electric field drives the particles along the z-direction and forces a clustering around the top of the helix windings.

\paragraph*{Potential Landscape}
The fundamental change of system properties for large $E$ is evident from the energies of individual minima. 
They can be seen in Fig. \ref{figure5}(c), again sorted by the number of singlets $s$.
Compared to the case $E=0$ shown in Fig. \ref{figure2}(d), the minima are now sorted into narrow \textquoteleft bands\textquoteright\ with little relative distance in energy.
The minima of each band contain the same number of singlets, doublets and so on.
These bands are formed because, for these large fields, the interaction between particles is only significant when they share a winding.
Therefore, for large $E$, the specific number of x-lets in a minimum has a greater impact on the energy than their relative ordering.

Furthermore, since the potential landscape for a dominant electric field is qualitatively described by the potential $V_E$, we can also estimate the stability of minima in this regime. 
Transitioning to another minimum requires a particle to change its winding, and therefore to move past the transition state at which the particle is located at the down most point of a winding.
Using Eq. \ref{eq:Ve} and neglecting the Coulomb interaction, this transition requires energy in the order of $\Delta E=2Er$.
In theory, these energy barriers can be made arbitrarily large.
The increased stability in large electric fields can also be used to tune a more common, temperature-induced, order-disorder transition.

\section{Generalization to Parameter Variations}
\label{Generalization}

We will now discuss the influence of parameter variations on the transition between amorphous and crystalline particle ordering. 
Since the clustering of particles at the top of the helix windings can always be enforced if $E$ is large enough and $r>0$, we will focus our discussion on arguments for the persistence of amorphous ordering for parameter variations in the absence of the external electric field.

\subsection{System Size}

In the previous section, we have observed the tendency of particles to avoid positions close to the center of the torus. 
This effect is a direct consequence of the small system size and can be suppressed by reducing the ratio of $r/R$. 
Here, we study the impact of the ratio $r/R$ on the system by increasing the winding number $M$ (and consequently increasing $R$ due to the relation $Mh=2\pi R$) while maintaining a constant filling factor $v=N/M$ and helix pitch $h$.
Note, that due to the large number of minima and the numerical challenge of finding \textit{all} of them, data for systems with $M>8$ were obtained from a random subset of minima and not every possible equilibrium configuration in the potential landscape for these specific parameters.
The number of minima $\eta$ considered for each subset has been chosen such that the total number of particles used for the statistics is about $\eta vM\sim 30000$.
To ensure that this statistics is representative for the system, we verified that repeated calculations with different random subsets give very similar results.

\begin{figure}
\includegraphics[width=\columnwidth]{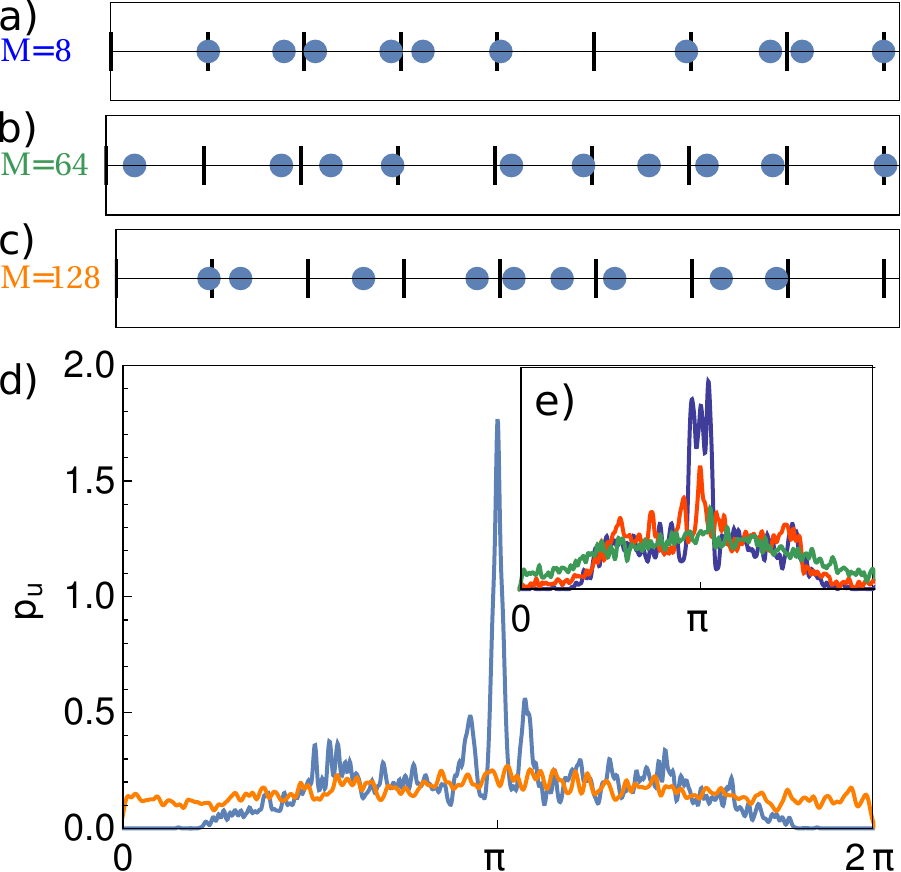}
\caption{(a)-(c) Particle positions (cutout of 8 windings) of equilibrium configurations in parametric coordinates for a constant filling $v=1.25$ and (a) $M=8$, (b) $M=64$, (a) $M=128$. (d) Probability density $p_u$ for finding a particle at a certain position in a winding for $M=8$ (blue) and $M=128$ (orange). The inset shows $p_u$ for $M=16,32,64$ (blue, red, green) respectively. \label{figure3}}
\end{figure}

The influence of increasing the winding number (and thereby increasing the torus radius) is analyzed in Fig. \ref{figure3}.
A general overview is given by Figs. \ref{figure3}(a)-(c)
which show the particle positions in parametric coordinates over a distance of eight windings for systems with a total of $M=[8,64,128]$ windings and a constant filling of $v=1.25$.
Fig. \ref{figure3}(a) represents a system with $M=8$, 
where all particle positions are close to the outermost part of the windings which are indicated by the vertical lines.
With increasing $M$ (Figs. \ref{figure3}(b,c)), this specific position dependence disappears such that in (c) the particle positions look almost random.

Further insight into the size-effects can be gained from the analysis in Fig. \ref{figure3}(d).
It shows the probability density $p_u$ for finding a particle at a specific position within a winding for a system with $M=8$ (blue) and a much larger setup with $M=128$ (orange).
The positions $0$ and $2\pi$ correspond to the positions closest to the center of the torus, whereas $\pi$ marks the outermost part of the winding (with a distance of $R+r$ from the torus center).
We can see that, for $M=8$, particles avoid positions close to the center of the torus.
In addition, there is a peak at $\pi$, indicating an increased likelihood of finding a particle at an outermost position. 
This peak is mostly caused by the singlets which - due to their lack of close neighbors - hardly deviate in terms of their position from the outermost part of the winding. 
In contrast, this dependence of $p_u$ on the position within a winding is hardly visible for $M=128$. 
For $M=128$, the system is a good approximation of the straight helix with $M,R\rightarrow\infty$ and a constant probability density of $1/2\pi$.
Interestingly, (except for the peak at $\pi$) the probability density $p_u$ for $M=8$ is of similar order in the range $[\pi/2, 3\pi/2]$, indicating that the particle positions for $M=8$ are far less predictable than the visualizations of Figs. \ref{figure2}(e)-(i) might suggest.

Fig. \ref{figure3}(d) shows only the cases of very small ($M=8$) and very large ($M=128$) systems.
A better understanding of the evolution of the probability distribution $p_u$ between these two regimes can be gained from the inset (e) of Fig. \ref{figure3}.
It shows the curves $p_u$ for different intermediate values of $M$.
An interesting observation is that for increasing $M$ the peak at $\pi$ broadens and then disappears, before any significant change in the `avoided' region around $0$ and $2\pi$ can be observed. 
Only after the peak at $\pi$ disappeared, does the probability for finding a particle close to $0$ or $2\pi$ increase. 
The reason for this is that when $M$ is increased, the forces of NNs begin to dominate over the combined interactions with all other particles. 
In general, the positions of singlets are more strongly affected than doublets or triplets since they, due to their lack of close neighbors, require less energy to slightly shift their position. 
Only for (much) larger $M$, the torus gets big enough such that the effect of particles avoiding positions close to the center of the torus becomes negligible. 
The described behavior implies that for increasing $r/R$ the particles deviate from the pattern for equilibrium particle positions described in the previous section. 

An increase of the winding number is therefore responsible for an increase of the disorder - at first due to the shift in singlet positions and for larger $M$ due to the loss of any preferred position within the windings. 
Specifically for the values of Fig. \ref{figure3}(d): when the winding number is increased from $M=8$ to $M=128$, the distribution of $p_{\Theta}$ changes from the values of Fig. \ref{figure4} ($\overline{p_{\Theta}}=0.42$ and $\sigma(p_{\Theta})=0.071$) to $\overline{p_{\Theta}}=0.24$ and $\sigma(p_{\Theta})=0.03$.
Since the amorphous particle order persists in the limiting case of the straight helix, we can assume that the described behavior is valid for a large parameter range of helical systems.

\subsection{Impact of the Helix Radius and the Filling Factor}

\begin{figure}
\includegraphics[width=\columnwidth]{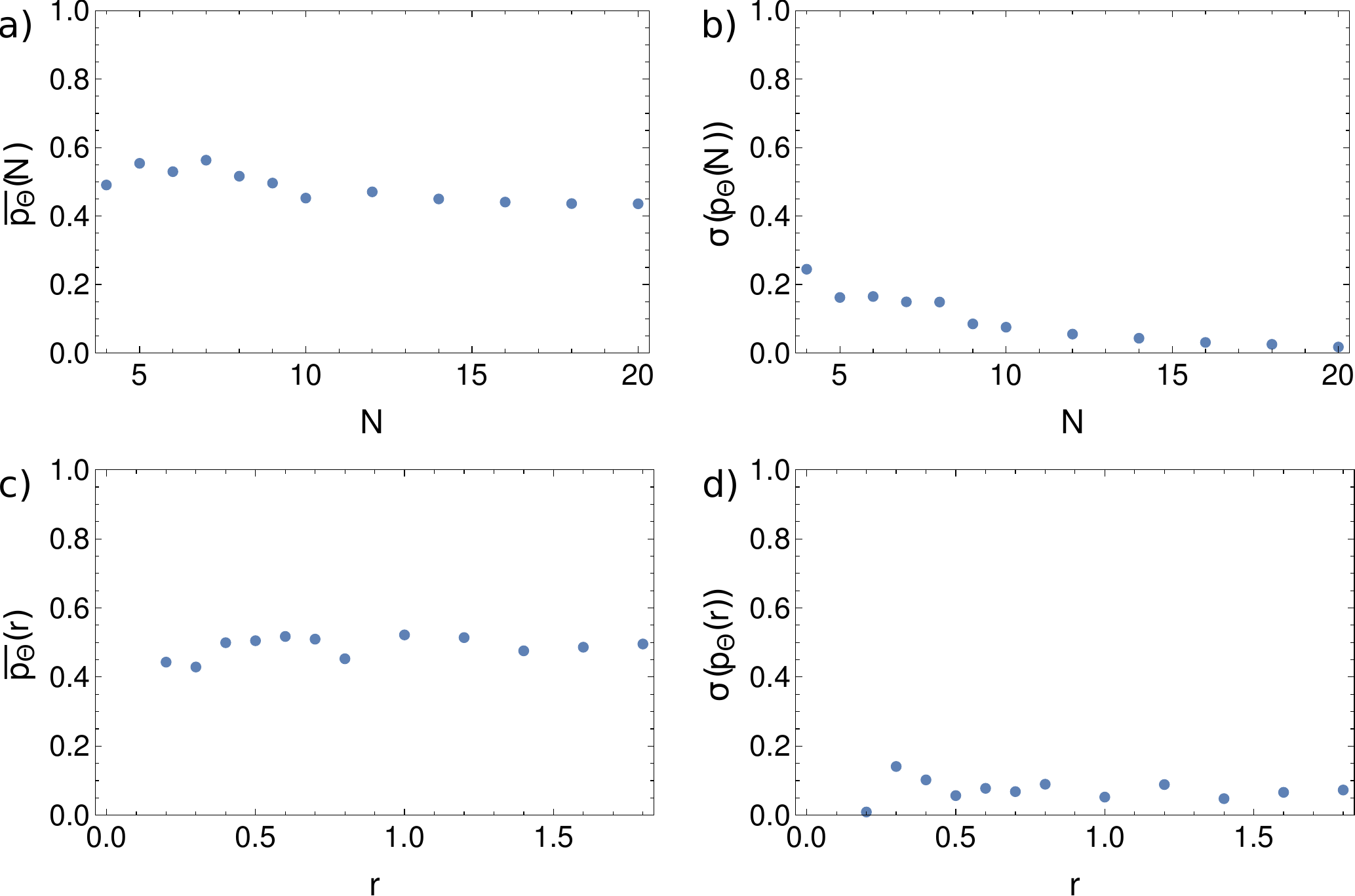}
\caption{The mean $\overline{p_{\Theta}}$ and variance $\sigma(p_{\Theta})$ of the order parameter for variations of (a)-(b) the particle number $N$, and (c)-(d) the helix radius $r$. All figures were obtained for $M=8$ and $E=0$. \label{FigureGeneralization}}
\end{figure}

\paragraph*{Filling Factor}
So far, we have explored the impact of the winding number on the properties of our helical setup, while keeping the filling factor $v$ and the helix radius $r$ constant. 
Now we examine the respective effects of the latter two parameters on the particle order for $E=0$. 
We investigate the filling factor $v=N/M$ by varying the particle number $N$ while maintaining a constant number of windings $M$ to prevent the occurrence of additional effects on the order parameter due to the system size. 
The mean and variance of the distribution $p_{\Theta}$ for different particle numbers are shown in Fig. \ref{FigureGeneralization}(a) and (b), where the particle number is varied in the range $4...20$. 
The mean value is approximately $\overline{p_{\Theta}}\sim0.5$, while the variance decreases with increasing $N$. 

The overall behavior of $p_{\Theta}$ with changes in the particle number can be explained by the effect of the changing particle density on the helix. 
It is in general unfavorable for a winding to have a much larger particle density than its surrounding windings. 
However, a larger particle density on the helix requires more particles within each winding, and consequently lower-order x-lets, such as singlets or doublets, are increasingly suppressed with increasing filling factor $v$.
Compare for example the case of $N=10$ to $N=20$:
In the former, the average equilibrium particle configuration contains about $\sim 2.8$ singlets ($\sim34\%$), whereas, for $N=20$, the average number of singlets per equilibrium configuration is about $\sim 1.4$ ($\sim18\%$). 
We also know from the discussion of the previous section, that the NN distances of particles that share a winding are very similar. 
Since our order parameter only requires these NN distances as input, slightly different particle distributions among the windings will (for large $N$) only have a small impact on the order parameter. 
Consequently, the variance of the distribution $p_{\Theta}$ decreases with increasing filling factor $v$.

A final remark is in order for the extreme cases of very low ($N\ll M$) and very large ($N\gg M$) filling factors. 
While the parameter range explored in Fig. \ref{FigureGeneralization}(a) and (b) does not have any adverse effect on the structural crossover, the disorder for $E=0$ can be affected in the limits $v\rightarrow0$ and $v\rightarrow\infty$. 
In case of $N\ll M$, when the particle density on the torus goes to zero, the possible equilibria will mostly consist of isolated particles with large distances to their nearest neighbors. 
In this case, the system is too scarcely populated to define any (useful) particle ordering. 
On the other hand, in the limit of $N\gg M$, the average particle distance will approach zero and the particle positions will almost exclusively be determined by their NNs. 
We assume that in this case, the equilibria for $E=0$ will consist of particles with crystalline ordering - although in small systems a slight position dependence of the lattice constant may appear due to the previously discussed size effects.

\paragraph*{Helix Radius}
The helix radius can be varied in the range $0\leq r\leq R$. 
When $r$ is varied within this range, we (again) obtain an almost constant mean value of $\overline{p_{\Theta}}\sim 0.5$ (see Fig. \ref{FigureGeneralization}(c)). 
While the corresponding variance $\sigma(p_{\Theta})$ is also approximately constant for most of the parameter range, it becomes zero below a critical helix radius (see Fig. \ref{FigureGeneralization}(d)).
This is because the number of equilibrium states can be tuned with the helix radius (specifically the ratio of $r/h$) and for low values of $r$, the system only possesses a single equilibrium state. 
Otherwise, the behavior of $p_{\Theta}$ for variations of $r$ is mainly determined by two competing effects: 
First, an increase in the helix radius increases the ratio of $r/R$ and thereby increases the size effect (and the order parameter) similar to the description above for variations of $M$. 
Second, at the same time, the helix radius also tunes the number of particles that can be stabilized within a single winding. 
From the discussion of Fig. \ref{figure4}, we know that equilibria with higher-order x-lets will be less ordered than e.g. those with a large number of singlets. 
Consequently, this second effect alone will lead to a decrease of the order parameter for increasing $r$. 
The values of Fig. \ref{FigureGeneralization}(c) and (d) are the result of the competition between those two effects.

\section{Summary and Conclusion}
\label{summary}

In this work, we considered a novel effective interaction arising from the confinement of Coulomb interacting particles to a 1D toroidal helix. 
The effective interaction allows for the existence of a plethora of stable equilibrium particle configurations. 
We investigated the properties of these equilibrium configurations in the presence of an external electric field and found a structural crossover that can be tuned by varying the electric field strength. 
For a vanishing electric field, we found a preference for amorphous particle ordering, whereas, in the regime where the electric field dominates over the Coulomb interaction, the particles cluster within the helix windings and order themselves in crystalline structures. 
Especially in the regime of low electric fields, the specific particle positions of the equilibria can depend on the helix geometry and the particle number. 
We therefore also explored the general effects occurring when these parameters are varied. 
While some parameter variations can influence the overall order of the equilibria, the amorphous ordering for $E=0$ persists for a large parameter range. 
Notable limits to this parameter range are low helix radii and both the upper and lower limits of the filling factor.

A natural continuation of this work consists in the investigation of the quantum mechanics of helically confined particle chains. 
Here, the question arises whether or not we obtain a fundamentally different behavior compared to the classical system. 
In case of equilibrium states, this concerns the structure of the eigenstates and the question for the existence of a quantum mechanical counterpart for the multitude of minima in the classical description. 
For the dynamics, it will be interesting to see if the intriguing phenomena found in the classical description \cite{zampetaki2013,zampetaki2015,zampetaki2015a,zampetaki2017} survive and to what extent they are modified. 
First steps towards a quantum mechanical description of particle chains in helical confinement were already taken in Refs. \cite{pedersen2016,pedersen2016a}.

Another promising direction for future works is the study of the optical properties of particles in a curved confinement. 
Helical 1D nanowires have already been proposed as terahertz antennas \cite{collier2018}.
Modeling the system as nonlinearly coupled oscillators, the observation of complex dynamics - such as higher harmonics generation or period-doubling under external driving - is very likely.
In this case, the electric field could be used to switch from a disordered state with a highly nontrivial optical response to a more ordered regime.

\begin{acknowledgments}
A. S. thanks Aritra K. Mukhopadhyay for a careful reading of the manuscript.
\end{acknowledgments}

\bibliography{orderDisorder_v36.bib}

\begin{thebibliography}{38}%
\makeatletter
\providecommand \@ifxundefined [1]{%
 \@ifx{#1\undefined}
}%
\providecommand \@ifnum [1]{%
 \ifnum #1\expandafter \@firstoftwo
 \else \expandafter \@secondoftwo
 \fi
}%
\providecommand \@ifx [1]{%
 \ifx #1\expandafter \@firstoftwo
 \else \expandafter \@secondoftwo
 \fi
}%
\providecommand \natexlab [1]{#1}%
\providecommand \enquote  [1]{``#1''}%
\providecommand \bibnamefont  [1]{#1}%
\providecommand \bibfnamefont [1]{#1}%
\providecommand \citenamefont [1]{#1}%
\providecommand \href@noop [0]{\@secondoftwo}%
\providecommand \href [0]{\begingroup \@sanitize@url \@href}%
\providecommand \@href[1]{\@@startlink{#1}\@@href}%
\providecommand \@@href[1]{\endgroup#1\@@endlink}%
\providecommand \@sanitize@url [0]{\catcode `\\12\catcode `\$12\catcode
  `\&12\catcode `\#12\catcode `\^12\catcode `\_12\catcode `\%12\relax}%
\providecommand \@@startlink[1]{}%
\providecommand \@@endlink[0]{}%
\providecommand \url  [0]{\begingroup\@sanitize@url \@url }%
\providecommand \@url [1]{\endgroup\@href {#1}{\urlprefix }}%
\providecommand \urlprefix  [0]{URL }%
\providecommand \Eprint [0]{\href }%
\providecommand \doibase [0]{http://dx.doi.org/}%
\providecommand \selectlanguage [0]{\@gobble}%
\providecommand \bibinfo  [0]{\@secondoftwo}%
\providecommand \bibfield  [0]{\@secondoftwo}%
\providecommand \translation [1]{[#1]}%
\providecommand \BibitemOpen [0]{}%
\providecommand \bibitemStop [0]{}%
\providecommand \bibitemNoStop [0]{.\EOS\space}%
\providecommand \EOS [0]{\spacefactor3000\relax}%
\providecommand \BibitemShut  [1]{\csname bibitem#1\endcsname}%
\let\auto@bib@innerbib\@empty
\bibitem [{\citenamefont {Lan}\ \emph {et~al.}(2011)\citenamefont {Lan},
  \citenamefont {Wang},\ and\ \citenamefont {Ren}}]{lan2011}%
  \BibitemOpen
  \bibfield  {author} {\bibinfo {author} {\bibfnamefont {Y.}~\bibnamefont
  {Lan}}, \bibinfo {author} {\bibfnamefont {Y.}~\bibnamefont {Wang}}, \ and\
  \bibinfo {author} {\bibfnamefont {Z.~F.}\ \bibnamefont {Ren}},\ }\href
  {\doibase 10.1080/00018732.2011.599963} {\bibfield  {journal} {\bibinfo
  {journal} {Advances in Physics}\ }\textbf {\bibinfo {volume} {60}},\ \bibinfo
  {pages} {553} (\bibinfo {year} {2011})}\BibitemShut {NoStop}%
\bibitem [{\citenamefont {Samuelson}\ \emph {et~al.}(2004)\citenamefont
  {Samuelson}, \citenamefont {Thelander}, \citenamefont {Bj{\"o}rk},
  \citenamefont {Borgstr{\"o}m}, \citenamefont {Deppert}, \citenamefont {Dick},
  \citenamefont {Hansen}, \citenamefont {M{\aa}rtensson}, \citenamefont
  {Panev}, \citenamefont {Persson}, \citenamefont {Seifert}, \citenamefont
  {Sk{\"o}ld}, \citenamefont {Larsson},\ and\ \citenamefont
  {Wallenberg}}]{samuelson2004}%
  \BibitemOpen
  \bibfield  {author} {\bibinfo {author} {\bibfnamefont {L.}~\bibnamefont
  {Samuelson}}, \bibinfo {author} {\bibfnamefont {C.}~\bibnamefont
  {Thelander}}, \bibinfo {author} {\bibfnamefont {M.}~\bibnamefont
  {Bj{\"o}rk}}, \bibinfo {author} {\bibfnamefont {M.}~\bibnamefont
  {Borgstr{\"o}m}}, \bibinfo {author} {\bibfnamefont {K.}~\bibnamefont
  {Deppert}}, \bibinfo {author} {\bibfnamefont {K.}~\bibnamefont {Dick}},
  \bibinfo {author} {\bibfnamefont {A.}~\bibnamefont {Hansen}}, \bibinfo
  {author} {\bibfnamefont {T.}~\bibnamefont {M{\aa}rtensson}}, \bibinfo
  {author} {\bibfnamefont {N.}~\bibnamefont {Panev}}, \bibinfo {author}
  {\bibfnamefont {A.}~\bibnamefont {Persson}}, \bibinfo {author} {\bibfnamefont
  {W.}~\bibnamefont {Seifert}}, \bibinfo {author} {\bibfnamefont
  {N.}~\bibnamefont {Sk{\"o}ld}}, \bibinfo {author} {\bibfnamefont
  {M.}~\bibnamefont {Larsson}}, \ and\ \bibinfo {author} {\bibfnamefont
  {L.}~\bibnamefont {Wallenberg}},\ }\href {\doibase
  10.1016/j.physe.2004.06.030} {\bibfield  {journal} {\bibinfo  {journal}
  {Physica E: Low-dimensional Systems and Nanostructures}\ }\textbf {\bibinfo
  {volume} {25}},\ \bibinfo {pages} {313} (\bibinfo {year} {2004})}\BibitemShut
  {NoStop}%
\bibitem [{\citenamefont {Avouris}\ and\ \citenamefont
  {Chen}(2006)}]{avouris2006}%
  \BibitemOpen
  \bibfield  {author} {\bibinfo {author} {\bibfnamefont {P.}~\bibnamefont
  {Avouris}}\ and\ \bibinfo {author} {\bibfnamefont {J.}~\bibnamefont {Chen}},\
  }\href {\doibase 10.1016/S1369-7021(06)71653-4} {\bibfield  {journal}
  {\bibinfo  {journal} {Materials Today}\ }\textbf {\bibinfo {volume} {9}},\
  \bibinfo {pages} {46} (\bibinfo {year} {2006})}\BibitemShut {NoStop}%
\bibitem [{\citenamefont {Hatton}\ \emph {et~al.}(2008)\citenamefont {Hatton},
  \citenamefont {Miller},\ and\ \citenamefont {Silva}}]{hatton2008}%
  \BibitemOpen
  \bibfield  {author} {\bibinfo {author} {\bibfnamefont {R.~A.}\ \bibnamefont
  {Hatton}}, \bibinfo {author} {\bibfnamefont {A.~J.}\ \bibnamefont {Miller}},
  \ and\ \bibinfo {author} {\bibfnamefont {S.~R.~P.}\ \bibnamefont {Silva}},\
  }\href {\doibase 10.1039/b713527k} {\bibfield  {journal} {\bibinfo  {journal}
  {Journal of Materials Chemistry}\ }\textbf {\bibinfo {volume} {18}},\
  \bibinfo {pages} {1183} (\bibinfo {year} {2008})}\BibitemShut {NoStop}%
\bibitem [{\citenamefont {Avouris}\ \emph {et~al.}(2008)\citenamefont
  {Avouris}, \citenamefont {Freitag},\ and\ \citenamefont
  {Perebeinos}}]{avouris2008}%
  \BibitemOpen
  \bibfield  {author} {\bibinfo {author} {\bibfnamefont {P.}~\bibnamefont
  {Avouris}}, \bibinfo {author} {\bibfnamefont {M.}~\bibnamefont {Freitag}}, \
  and\ \bibinfo {author} {\bibfnamefont {V.}~\bibnamefont {Perebeinos}},\
  }\href {\doibase 10.1038/nphoton.2008.94} {\bibfield  {journal} {\bibinfo
  {journal} {Nature Photonics}\ }\textbf {\bibinfo {volume} {2}},\ \bibinfo
  {pages} {341} (\bibinfo {year} {2008})}\BibitemShut {NoStop}%
\bibitem [{\citenamefont {Ijima}\ and\ \citenamefont
  {Ichihashi}(1993)}]{ijima1993}%
  \BibitemOpen
  \bibfield  {author} {\bibinfo {author} {\bibfnamefont {S.}~\bibnamefont
  {Ijima}}\ and\ \bibinfo {author} {\bibfnamefont {T.}~\bibnamefont
  {Ichihashi}},\ }\href {\doibase 10.1038/363603a0} {\bibfield  {journal}
  {\bibinfo  {journal} {Nature}\ }\textbf {\bibinfo {volume} {363}},\ \bibinfo
  {pages} {603} (\bibinfo {year} {1993})}\BibitemShut {NoStop}%
\bibitem [{\citenamefont {Shaikjee}\ and\ \citenamefont
  {Coville}(2012)}]{shaikjee2012}%
  \BibitemOpen
  \bibfield  {author} {\bibinfo {author} {\bibfnamefont {A.}~\bibnamefont
  {Shaikjee}}\ and\ \bibinfo {author} {\bibfnamefont {N.~J.}\ \bibnamefont
  {Coville}},\ }\href {\doibase 10.1016/j.jare.2011.05.007} {\bibfield
  {journal} {\bibinfo  {journal} {Journal of Advanced Research}\ }\textbf
  {\bibinfo {volume} {3}},\ \bibinfo {pages} {195} (\bibinfo {year}
  {2012})}\BibitemShut {NoStop}%
\bibitem [{\citenamefont {Lau}\ \emph {et~al.}(2006)\citenamefont {Lau},
  \citenamefont {Lu},\ and\ \citenamefont {Hui}}]{lau2006}%
  \BibitemOpen
  \bibfield  {author} {\bibinfo {author} {\bibfnamefont {K.~T.}\ \bibnamefont
  {Lau}}, \bibinfo {author} {\bibfnamefont {M.}~\bibnamefont {Lu}}, \ and\
  \bibinfo {author} {\bibfnamefont {D.}~\bibnamefont {Hui}},\ }\href {\doibase
  10.1016/j.compositesb.2006.02.008} {\bibfield  {journal} {\bibinfo  {journal}
  {Composites Part B: Engineering}\ }\textbf {\bibinfo {volume} {37}},\
  \bibinfo {pages} {437} (\bibinfo {year} {2006})}\BibitemShut {NoStop}%
\bibitem [{\citenamefont {{Garcia-Fernandez}}\ \emph
  {et~al.}(2011)\citenamefont {{Garcia-Fernandez}}, \citenamefont {Alt},
  \citenamefont {Bruse}, \citenamefont {Dan}, \citenamefont {Karapetyan},
  \citenamefont {Rehband}, \citenamefont {Stiebeiner}, \citenamefont
  {Wiedemann}, \citenamefont {Meschede},\ and\ \citenamefont
  {Rauschenbeutel}}]{garcia-fernandez2011}%
  \BibitemOpen
  \bibfield  {author} {\bibinfo {author} {\bibfnamefont {R.}~\bibnamefont
  {{Garcia-Fernandez}}}, \bibinfo {author} {\bibfnamefont {W.}~\bibnamefont
  {Alt}}, \bibinfo {author} {\bibfnamefont {F.}~\bibnamefont {Bruse}}, \bibinfo
  {author} {\bibfnamefont {C.}~\bibnamefont {Dan}}, \bibinfo {author}
  {\bibfnamefont {K.}~\bibnamefont {Karapetyan}}, \bibinfo {author}
  {\bibfnamefont {O.}~\bibnamefont {Rehband}}, \bibinfo {author} {\bibfnamefont
  {A.}~\bibnamefont {Stiebeiner}}, \bibinfo {author} {\bibfnamefont
  {U.}~\bibnamefont {Wiedemann}}, \bibinfo {author} {\bibfnamefont
  {D.}~\bibnamefont {Meschede}}, \ and\ \bibinfo {author} {\bibfnamefont
  {A.}~\bibnamefont {Rauschenbeutel}},\ }\href {\doibase
  10.1007/s00340-011-4730-x} {\bibfield  {journal} {\bibinfo  {journal}
  {Applied Physics B}\ }\textbf {\bibinfo {volume} {105}},\ \bibinfo {pages}
  {3} (\bibinfo {year} {2011})}\BibitemShut {NoStop}%
\bibitem [{\citenamefont {Vetsch}\ \emph {et~al.}(2012)\citenamefont {Vetsch},
  \citenamefont {Dawkins}, \citenamefont {Mitsch}, \citenamefont {Reitz},
  \citenamefont {Schneeweiss},\ and\ \citenamefont
  {Rauschenbeutel}}]{vetsch2012}%
  \BibitemOpen
  \bibfield  {author} {\bibinfo {author} {\bibfnamefont {E.}~\bibnamefont
  {Vetsch}}, \bibinfo {author} {\bibfnamefont {S.~T.}\ \bibnamefont {Dawkins}},
  \bibinfo {author} {\bibfnamefont {R.}~\bibnamefont {Mitsch}}, \bibinfo
  {author} {\bibfnamefont {D.}~\bibnamefont {Reitz}}, \bibinfo {author}
  {\bibfnamefont {P.}~\bibnamefont {Schneeweiss}}, \ and\ \bibinfo {author}
  {\bibfnamefont {A.}~\bibnamefont {Rauschenbeutel}},\ }\href {\doibase
  10.1109/JSTQE.2012.2196025} {\bibfield  {journal} {\bibinfo  {journal} {IEEE
  Journal of Selected Topics in Quantum Electronics}\ }\textbf {\bibinfo
  {volume} {18}},\ \bibinfo {pages} {1763} (\bibinfo {year}
  {2012})}\BibitemShut {NoStop}%
\bibitem [{\citenamefont {Reitz}\ and\ \citenamefont
  {Rauschenbeutel}(2012)}]{reitz2012}%
  \BibitemOpen
  \bibfield  {author} {\bibinfo {author} {\bibfnamefont {D.}~\bibnamefont
  {Reitz}}\ and\ \bibinfo {author} {\bibfnamefont {A.}~\bibnamefont
  {Rauschenbeutel}},\ }\href {\doibase 10.1016/j.optcom.2012.06.034} {\bibfield
   {journal} {\bibinfo  {journal} {Optics Communications}\ }\textbf {\bibinfo
  {volume} {285}},\ \bibinfo {pages} {4705} (\bibinfo {year}
  {2012})}\BibitemShut {NoStop}%
\bibitem [{\citenamefont {Vetsch}\ \emph {et~al.}(2010)\citenamefont {Vetsch},
  \citenamefont {Reitz}, \citenamefont {Sagu{\'e}}, \citenamefont {Schmidt},
  \citenamefont {Dawkins},\ and\ \citenamefont {Rauschenbeutel}}]{vetsch2010}%
  \BibitemOpen
  \bibfield  {author} {\bibinfo {author} {\bibfnamefont {E.}~\bibnamefont
  {Vetsch}}, \bibinfo {author} {\bibfnamefont {D.}~\bibnamefont {Reitz}},
  \bibinfo {author} {\bibfnamefont {G.}~\bibnamefont {Sagu{\'e}}}, \bibinfo
  {author} {\bibfnamefont {R.}~\bibnamefont {Schmidt}}, \bibinfo {author}
  {\bibfnamefont {S.~T.}\ \bibnamefont {Dawkins}}, \ and\ \bibinfo {author}
  {\bibfnamefont {A.}~\bibnamefont {Rauschenbeutel}},\ }\href {\doibase
  10.1103/PhysRevLett.104.203603} {\bibfield  {journal} {\bibinfo  {journal}
  {Physical Review Letters}\ }\textbf {\bibinfo {volume} {104}},\ \bibinfo
  {pages} {203603} (\bibinfo {year} {2010})}\BibitemShut {NoStop}%
\bibitem [{\citenamefont {Tran}\ \emph {et~al.}(2000)\citenamefont {Tran},
  \citenamefont {Alavi},\ and\ \citenamefont {Gruner}}]{tran2000}%
  \BibitemOpen
  \bibfield  {author} {\bibinfo {author} {\bibfnamefont {P.}~\bibnamefont
  {Tran}}, \bibinfo {author} {\bibfnamefont {B.}~\bibnamefont {Alavi}}, \ and\
  \bibinfo {author} {\bibfnamefont {G.}~\bibnamefont {Gruner}},\ }\href@noop {}
  {\bibfield  {journal} {\bibinfo  {journal} {Physical Review Letters}\
  }\textbf {\bibinfo {volume} {85}},\ \bibinfo {pages} {4} (\bibinfo {year}
  {2000})}\BibitemShut {NoStop}%
\bibitem [{\citenamefont {Yu}\ and\ \citenamefont {Song}(2001)}]{yu2001}%
  \BibitemOpen
  \bibfield  {author} {\bibinfo {author} {\bibfnamefont {Z.}~\bibnamefont
  {Yu}}\ and\ \bibinfo {author} {\bibfnamefont {X.}~\bibnamefont {Song}},\
  }\href {\doibase 10.1103/PhysRevLett.86.6018} {\bibfield  {journal} {\bibinfo
   {journal} {Physical Review Letters}\ }\textbf {\bibinfo {volume} {86}},\
  \bibinfo {pages} {6018} (\bibinfo {year} {2001})}\BibitemShut {NoStop}%
\bibitem [{\citenamefont {Marko}\ and\ \citenamefont
  {Siggia}(1994)}]{marko1994}%
  \BibitemOpen
  \bibfield  {author} {\bibinfo {author} {\bibfnamefont {J.~F.}\ \bibnamefont
  {Marko}}\ and\ \bibinfo {author} {\bibfnamefont {E.~D.}\ \bibnamefont
  {Siggia}},\ }\href {\doibase 10.1021/ma00082a015} {\bibfield  {journal}
  {\bibinfo  {journal} {Macromolecules}\ }\textbf {\bibinfo {volume} {27}},\
  \bibinfo {pages} {981} (\bibinfo {year} {1994})}\BibitemShut {NoStop}%
\bibitem [{\citenamefont {Kornyshev}\ \emph {et~al.}(2007)\citenamefont
  {Kornyshev}, \citenamefont {Lee}, \citenamefont {Leikin},\ and\ \citenamefont
  {Wynveen}}]{kornyshev2007}%
  \BibitemOpen
  \bibfield  {author} {\bibinfo {author} {\bibfnamefont {A.~A.}\ \bibnamefont
  {Kornyshev}}, \bibinfo {author} {\bibfnamefont {D.~J.}\ \bibnamefont {Lee}},
  \bibinfo {author} {\bibfnamefont {S.}~\bibnamefont {Leikin}}, \ and\ \bibinfo
  {author} {\bibfnamefont {A.}~\bibnamefont {Wynveen}},\ }\href {\doibase
  10.1103/RevModPhys.79.943} {\bibfield  {journal} {\bibinfo  {journal}
  {Reviews of Modern Physics}\ }\textbf {\bibinfo {volume} {79}},\ \bibinfo
  {pages} {943} (\bibinfo {year} {2007})}\BibitemShut {NoStop}%
\bibitem [{\citenamefont {Prinz}\ \emph {et~al.}(2000)\citenamefont {Prinz},
  \citenamefont {Seleznev}, \citenamefont {Gutakovsky}, \citenamefont
  {Chehovskiy}, \citenamefont {Preobrazhenskii}, \citenamefont {Putyato},\ and\
  \citenamefont {Gavrilova}}]{prinz2000}%
  \BibitemOpen
  \bibfield  {author} {\bibinfo {author} {\bibfnamefont {V.}~\bibnamefont
  {Prinz}}, \bibinfo {author} {\bibfnamefont {V.}~\bibnamefont {Seleznev}},
  \bibinfo {author} {\bibfnamefont {A.}~\bibnamefont {Gutakovsky}}, \bibinfo
  {author} {\bibfnamefont {A.}~\bibnamefont {Chehovskiy}}, \bibinfo {author}
  {\bibfnamefont {V.}~\bibnamefont {Preobrazhenskii}}, \bibinfo {author}
  {\bibfnamefont {M.}~\bibnamefont {Putyato}}, \ and\ \bibinfo {author}
  {\bibfnamefont {T.}~\bibnamefont {Gavrilova}},\ }\href {\doibase
  10.1016/S1386-9477(99)00249-0} {\bibfield  {journal} {\bibinfo  {journal}
  {Physica E: Low-dimensional Systems and Nanostructures}\ }\textbf {\bibinfo
  {volume} {6}},\ \bibinfo {pages} {828} (\bibinfo {year} {2000})}\BibitemShut
  {NoStop}%
\bibitem [{\citenamefont {Schmidt}\ and\ \citenamefont
  {Eberl}(2001)}]{schmidt2001}%
  \BibitemOpen
  \bibfield  {author} {\bibinfo {author} {\bibfnamefont {O.~G.}\ \bibnamefont
  {Schmidt}}\ and\ \bibinfo {author} {\bibfnamefont {K.}~\bibnamefont
  {Eberl}},\ }\href {\doibase 10.1038/35065525} {\bibfield  {journal} {\bibinfo
   {journal} {Nature}\ }\textbf {\bibinfo {volume} {410}},\ \bibinfo {pages}
  {168} (\bibinfo {year} {2001})}\BibitemShut {NoStop}%
\bibitem [{\citenamefont {Bhattacharya}(2007)}]{bhattacharya2007}%
  \BibitemOpen
  \bibfield  {author} {\bibinfo {author} {\bibfnamefont {M.}~\bibnamefont
  {Bhattacharya}},\ }\href {\doibase 10.1016/j.optcom.2007.07.008} {\bibfield
  {journal} {\bibinfo  {journal} {Optics Communications}\ }\textbf {\bibinfo
  {volume} {279}},\ \bibinfo {pages} {219} (\bibinfo {year}
  {2007})}\BibitemShut {NoStop}%
\bibitem [{\citenamefont {Ivanov}\ \emph {et~al.}(1994)\citenamefont {Ivanov},
  \citenamefont {Nagy}, \citenamefont {Lambin}, \citenamefont {Lucas},
  \citenamefont {Zhang}, \citenamefont {Zhang}, \citenamefont {Bernaerts},
  \citenamefont {Van~Tendeloo}, \citenamefont {Amelinckx},\ and\ \citenamefont
  {Van~Landuyt}}]{ivanov1994}%
  \BibitemOpen
  \bibfield  {author} {\bibinfo {author} {\bibfnamefont {V.}~\bibnamefont
  {Ivanov}}, \bibinfo {author} {\bibfnamefont {J.}~\bibnamefont {Nagy}},
  \bibinfo {author} {\bibfnamefont {P.}~\bibnamefont {Lambin}}, \bibinfo
  {author} {\bibfnamefont {A.}~\bibnamefont {Lucas}}, \bibinfo {author}
  {\bibfnamefont {X.~B.}\ \bibnamefont {Zhang}}, \bibinfo {author}
  {\bibfnamefont {X.~F.}\ \bibnamefont {Zhang}}, \bibinfo {author}
  {\bibfnamefont {D.}~\bibnamefont {Bernaerts}}, \bibinfo {author}
  {\bibfnamefont {G.}~\bibnamefont {Van~Tendeloo}}, \bibinfo {author}
  {\bibfnamefont {S.}~\bibnamefont {Amelinckx}}, \ and\ \bibinfo {author}
  {\bibfnamefont {J.}~\bibnamefont {Van~Landuyt}},\ }\href {\doibase
  10.1016/0009-2614(94)00467-6} {\bibfield  {journal} {\bibinfo  {journal}
  {Chemical Physics Letters}\ }\textbf {\bibinfo {volume} {223}},\ \bibinfo
  {pages} {329} (\bibinfo {year} {1994})}\BibitemShut {NoStop}%
\bibitem [{\citenamefont {Drude}(1906)}]{drude1906}%
  \BibitemOpen
  \bibfield  {author} {\bibinfo {author} {\bibfnamefont {P.}~\bibnamefont
  {Drude}},\ }\href@noop {} {\emph {\bibinfo {title} {Lehrbuch Der
  {{Optik}}}}},\ \bibinfo {edition} {2nd}\ ed.\ (\bibinfo  {publisher}
  {{Hirzel}},\ \bibinfo {address} {{Leipzig}},\ \bibinfo {year}
  {1906})\BibitemShut {NoStop}%
\bibitem [{\citenamefont {Maki}\ and\ \citenamefont
  {Persoons}(1996)}]{maki1996}%
  \BibitemOpen
  \bibfield  {author} {\bibinfo {author} {\bibfnamefont {J.~J.}\ \bibnamefont
  {Maki}}\ and\ \bibinfo {author} {\bibfnamefont {A.}~\bibnamefont
  {Persoons}},\ }\href {\doibase 10.1063/1.471679} {\bibfield  {journal}
  {\bibinfo  {journal} {The Journal of Chemical Physics}\ }\textbf {\bibinfo
  {volume} {104}},\ \bibinfo {pages} {9340} (\bibinfo {year}
  {1996})}\BibitemShut {NoStop}%
\bibitem [{\citenamefont {Schmelcher}(2011)}]{schmelcher2011}%
  \BibitemOpen
  \bibfield  {author} {\bibinfo {author} {\bibfnamefont {P.}~\bibnamefont
  {Schmelcher}},\ }\href {\doibase 10.1209/0295-5075/95/50005} {\bibfield
  {journal} {\bibinfo  {journal} {Europhysics Letters}\ }\textbf {\bibinfo
  {volume} {95}},\ \bibinfo {pages} {50005} (\bibinfo {year}
  {2011})}\BibitemShut {NoStop}%
\bibitem [{\citenamefont {Kibis}(1992)}]{kibis1992}%
  \BibitemOpen
  \bibfield  {author} {\bibinfo {author} {\bibfnamefont {O.}~\bibnamefont
  {Kibis}},\ }\href {\doibase 10.1016/0375-9601(92)90730-A} {\bibfield
  {journal} {\bibinfo  {journal} {Physics Letters A}\ }\textbf {\bibinfo
  {volume} {166}},\ \bibinfo {pages} {393} (\bibinfo {year}
  {1992})}\BibitemShut {NoStop}%
\bibitem [{\citenamefont {Pedersen}\ \emph {et~al.}(2014)\citenamefont
  {Pedersen}, \citenamefont {Fedorov}, \citenamefont {Jensen},\ and\
  \citenamefont {Zinner}}]{pedersen2014}%
  \BibitemOpen
  \bibfield  {author} {\bibinfo {author} {\bibfnamefont {J.~K.}\ \bibnamefont
  {Pedersen}}, \bibinfo {author} {\bibfnamefont {D.~V.}\ \bibnamefont
  {Fedorov}}, \bibinfo {author} {\bibfnamefont {A.~S.}\ \bibnamefont {Jensen}},
  \ and\ \bibinfo {author} {\bibfnamefont {N.~T.}\ \bibnamefont {Zinner}},\
  }\href {\doibase 10.1088/0953-4075/47/16/165103} {\bibfield  {journal}
  {\bibinfo  {journal} {Journal of Physics B: Atomic, Molecular and Optical
  Physics}\ }\textbf {\bibinfo {volume} {47}},\ \bibinfo {pages} {165103}
  (\bibinfo {year} {2014})}\BibitemShut {NoStop}%
\bibitem [{\citenamefont {Zampetaki}\ \emph {et~al.}(2013)\citenamefont
  {Zampetaki}, \citenamefont {Stockhofe}, \citenamefont {Kr{\"o}nke},\ and\
  \citenamefont {Schmelcher}}]{zampetaki2013}%
  \BibitemOpen
  \bibfield  {author} {\bibinfo {author} {\bibfnamefont {A.~V.}\ \bibnamefont
  {Zampetaki}}, \bibinfo {author} {\bibfnamefont {J.}~\bibnamefont
  {Stockhofe}}, \bibinfo {author} {\bibfnamefont {S.}~\bibnamefont
  {Kr{\"o}nke}}, \ and\ \bibinfo {author} {\bibfnamefont {P.}~\bibnamefont
  {Schmelcher}},\ }\href {\doibase 10.1103/PhysRevE.88.043202} {\bibfield
  {journal} {\bibinfo  {journal} {Physical Review E}\ }\textbf {\bibinfo
  {volume} {88}},\ \bibinfo {pages} {043202} (\bibinfo {year}
  {2013})}\BibitemShut {NoStop}%
\bibitem [{\citenamefont {Zampetaki}\ \emph
  {et~al.}(2015{\natexlab{a}})\citenamefont {Zampetaki}, \citenamefont
  {Stockhofe},\ and\ \citenamefont {Schmelcher}}]{zampetaki2015}%
  \BibitemOpen
  \bibfield  {author} {\bibinfo {author} {\bibfnamefont {A.~V.}\ \bibnamefont
  {Zampetaki}}, \bibinfo {author} {\bibfnamefont {J.}~\bibnamefont
  {Stockhofe}}, \ and\ \bibinfo {author} {\bibfnamefont {P.}~\bibnamefont
  {Schmelcher}},\ }\href {\doibase 10.1103/PhysRevE.92.042905} {\bibfield
  {journal} {\bibinfo  {journal} {Physical Review E}\ }\textbf {\bibinfo
  {volume} {92}},\ \bibinfo {pages} {042905} (\bibinfo {year}
  {2015}{\natexlab{a}})}\BibitemShut {NoStop}%
\bibitem [{\citenamefont {Zampetaki}\ \emph
  {et~al.}(2015{\natexlab{b}})\citenamefont {Zampetaki}, \citenamefont
  {Stockhofe},\ and\ \citenamefont {Schmelcher}}]{zampetaki2015a}%
  \BibitemOpen
  \bibfield  {author} {\bibinfo {author} {\bibfnamefont {A.~V.}\ \bibnamefont
  {Zampetaki}}, \bibinfo {author} {\bibfnamefont {J.}~\bibnamefont
  {Stockhofe}}, \ and\ \bibinfo {author} {\bibfnamefont {P.}~\bibnamefont
  {Schmelcher}},\ }\href {\doibase 10.1103/PhysRevA.91.023409} {\bibfield
  {journal} {\bibinfo  {journal} {Physical Review A}\ }\textbf {\bibinfo
  {volume} {91}},\ \bibinfo {pages} {023409} (\bibinfo {year}
  {2015}{\natexlab{b}})}\BibitemShut {NoStop}%
\bibitem [{\citenamefont {Zampetaki}\ \emph {et~al.}(2017)\citenamefont
  {Zampetaki}, \citenamefont {Stockhofe},\ and\ \citenamefont
  {Schmelcher}}]{zampetaki2017}%
  \BibitemOpen
  \bibfield  {author} {\bibinfo {author} {\bibfnamefont {A.~V.}\ \bibnamefont
  {Zampetaki}}, \bibinfo {author} {\bibfnamefont {J.}~\bibnamefont
  {Stockhofe}}, \ and\ \bibinfo {author} {\bibfnamefont {P.}~\bibnamefont
  {Schmelcher}},\ }\href {\doibase 10.1103/PhysRevE.95.022205} {\bibfield
  {journal} {\bibinfo  {journal} {Physical Review E}\ }\textbf {\bibinfo
  {volume} {95}},\ \bibinfo {pages} {022205} (\bibinfo {year}
  {2017})}\BibitemShut {NoStop}%
\bibitem [{\citenamefont {Zampetaki}\ \emph {et~al.}(2018)\citenamefont
  {Zampetaki}, \citenamefont {Stockhofe},\ and\ \citenamefont
  {Schmelcher}}]{zampetaki2018}%
  \BibitemOpen
  \bibfield  {author} {\bibinfo {author} {\bibfnamefont {A.~V.}\ \bibnamefont
  {Zampetaki}}, \bibinfo {author} {\bibfnamefont {J.}~\bibnamefont
  {Stockhofe}}, \ and\ \bibinfo {author} {\bibfnamefont {P.}~\bibnamefont
  {Schmelcher}},\ }\href {\doibase 10.1103/PhysRevE.97.042503} {\bibfield
  {journal} {\bibinfo  {journal} {Physical Review E}\ }\textbf {\bibinfo
  {volume} {97}},\ \bibinfo {pages} {042503} (\bibinfo {year}
  {2018})}\BibitemShut {NoStop}%
\bibitem [{\citenamefont {Plettenberg}\ \emph {et~al.}(2017)\citenamefont
  {Plettenberg}, \citenamefont {Stockhofe}, \citenamefont {Zampetaki},\ and\
  \citenamefont {Schmelcher}}]{plettenberg2017}%
  \BibitemOpen
  \bibfield  {author} {\bibinfo {author} {\bibfnamefont {J.}~\bibnamefont
  {Plettenberg}}, \bibinfo {author} {\bibfnamefont {J.}~\bibnamefont
  {Stockhofe}}, \bibinfo {author} {\bibfnamefont {A.~V.}\ \bibnamefont
  {Zampetaki}}, \ and\ \bibinfo {author} {\bibfnamefont {P.}~\bibnamefont
  {Schmelcher}},\ }\href {\doibase 10.1103/PhysRevE.95.012213} {\bibfield
  {journal} {\bibinfo  {journal} {Physical Review E}\ }\textbf {\bibinfo
  {volume} {95}},\ \bibinfo {pages} {012213} (\bibinfo {year}
  {2017})}\BibitemShut {NoStop}%
\bibitem [{\citenamefont {Pedersen}\ \emph
  {et~al.}(2016{\natexlab{a}})\citenamefont {Pedersen}, \citenamefont
  {Fedorov}, \citenamefont {Jensen},\ and\ \citenamefont
  {Zinner}}]{pedersen2016a}%
  \BibitemOpen
  \bibfield  {author} {\bibinfo {author} {\bibfnamefont {J.~K.}\ \bibnamefont
  {Pedersen}}, \bibinfo {author} {\bibfnamefont {D.~V.}\ \bibnamefont
  {Fedorov}}, \bibinfo {author} {\bibfnamefont {A.~S.}\ \bibnamefont {Jensen}},
  \ and\ \bibinfo {author} {\bibfnamefont {N.~T.}\ \bibnamefont {Zinner}},\
  }\href {\doibase 10.1080/09500340.2015.1116634} {\bibfield  {journal}
  {\bibinfo  {journal} {Journal of Modern Optics}\ }\textbf {\bibinfo {volume}
  {63}},\ \bibinfo {pages} {1814} (\bibinfo {year}
  {2016}{\natexlab{a}})}\BibitemShut {NoStop}%
\bibitem [{\citenamefont {Pedersen}\ \emph
  {et~al.}(2016{\natexlab{b}})\citenamefont {Pedersen}, \citenamefont
  {Fedorov}, \citenamefont {Jensen},\ and\ \citenamefont
  {Zinner}}]{pedersen2016}%
  \BibitemOpen
  \bibfield  {author} {\bibinfo {author} {\bibfnamefont {J.~K.}\ \bibnamefont
  {Pedersen}}, \bibinfo {author} {\bibfnamefont {D.~V.}\ \bibnamefont
  {Fedorov}}, \bibinfo {author} {\bibfnamefont {A.~S.}\ \bibnamefont {Jensen}},
  \ and\ \bibinfo {author} {\bibfnamefont {N.~T.}\ \bibnamefont {Zinner}},\
  }\href {\doibase 10.1080/09500340.2015.1116634} {\bibfield  {journal}
  {\bibinfo  {journal} {Journal of Modern Optics}\ }\textbf {\bibinfo {volume}
  {63}},\ \bibinfo {pages} {1814} (\bibinfo {year}
  {2016}{\natexlab{b}})}\BibitemShut {NoStop}%
\bibitem [{\citenamefont {Scales}(1985)}]{scales1985}%
  \BibitemOpen
  \bibfield  {author} {\bibinfo {author} {\bibfnamefont {L.~E.}\ \bibnamefont
  {Scales}},\ }\href@noop {} {\emph {\bibinfo {title} {Introduction to
  {{Non}}-{{Linear Optimization}}}}}\ (\bibinfo  {publisher} {{Springer, New
  York, NY}},\ \bibinfo {year} {1985})\BibitemShut {NoStop}%
\bibitem [{\citenamefont {Nocedal}\ and\ \citenamefont
  {Wright}(2006)}]{nocedal2006}%
  \BibitemOpen
  \bibfield  {author} {\bibinfo {author} {\bibfnamefont {J.}~\bibnamefont
  {Nocedal}}\ and\ \bibinfo {author} {\bibfnamefont {S.~J.}\ \bibnamefont
  {Wright}},\ }\href@noop {} {\emph {\bibinfo {title} {Numerical
  {{Optimization}}}}}\ (\bibinfo  {publisher} {{Springer, New York, NY}},\
  \bibinfo {year} {2006})\BibitemShut {NoStop}%
\bibitem [{\citenamefont {Boyd}\ and\ \citenamefont
  {Vandenberghe}(2004)}]{boyd2004}%
  \BibitemOpen
  \bibfield  {author} {\bibinfo {author} {\bibfnamefont {S.~P.}\ \bibnamefont
  {Boyd}}\ and\ \bibinfo {author} {\bibfnamefont {L.}~\bibnamefont
  {Vandenberghe}},\ }\href@noop {} {\emph {\bibinfo {title} {Convex
  Optimization}}}\ (\bibinfo  {publisher} {{Cambridge University Press}},\
  \bibinfo {address} {{Cambridge}},\ \bibinfo {year} {2004})\BibitemShut
  {NoStop}%
\bibitem [{Note1()}]{Note1}%
  \BibitemOpen
  \bibinfo {note} {Windings are separated at positions closest to the center of
  the torus, i.e. odd multiples of $\pi $ in parametric
  coordinates}\BibitemShut {NoStop}%
\bibitem [{\citenamefont {Collier}\ \emph {et~al.}(2018)\citenamefont
  {Collier}, \citenamefont {Alexeev}, \citenamefont {Downing}, \citenamefont
  {Kibis},\ and\ \citenamefont {Portnoi}}]{collier2018}%
  \BibitemOpen
  \bibfield  {author} {\bibinfo {author} {\bibfnamefont {T.~P.}\ \bibnamefont
  {Collier}}, \bibinfo {author} {\bibfnamefont {A.~M.}\ \bibnamefont
  {Alexeev}}, \bibinfo {author} {\bibfnamefont {C.~A.}\ \bibnamefont
  {Downing}}, \bibinfo {author} {\bibfnamefont {O.~V.}\ \bibnamefont {Kibis}},
  \ and\ \bibinfo {author} {\bibfnamefont {M.~E.}\ \bibnamefont {Portnoi}},\
  }\href {\doibase 10.1134/S1063782618140075} {\bibfield  {journal} {\bibinfo
  {journal} {Semiconductors}\ }\textbf {\bibinfo {volume} {52}},\ \bibinfo
  {pages} {1813} (\bibinfo {year} {2018})}\BibitemShut {NoStop}%
\end{thebibliography}%
\end{document}